\documentclass[runningheads]{llncs}
\usepackage{graphicx}
\usepackage{pifont}
\usepackage{tikz}
\usepackage{comment}
\usepackage{amsmath,amssymb} 
\usepackage{color}
\usepackage{multirow}
\usepackage{makecell}
\usepackage[colorlinks, linkcolor=red, anchorcolor=yellow, citecolor=green]{hyperref}
\usepackage{lipsum}

\newcommand\blfootnote[1]{%
	\begingroup
	\renewcommand\thefootnote{}\footnote{#1}%
	\addtocounter{footnote}{-1}%
	\endgroup
}
\newcommand{\etal}{\textit{et al}. }
\newcommand{\ie}{\textit{i}.\textit{e}., }

\begin{document}
	\pagestyle{headings}
	\mainmatter
	\def\ECCVSubNumber{3}  
	
	\title{UDC-UNet: Under-Display Camera Image Restoration via U-shape Dynamic Network} 
	\titlerunning{UDC-UNet}
	\author{Xina Liu\inst{1,2}$^*$ \and Jinfan Hu\inst{1,2}$^*$
		\and Xiangyu Chen\inst{1,3,4} \and Chao Dong\inst{1,4}$^\dagger$}
	
	\authorrunning{Xina Liu, Jinfan Hu, Xiangyu Chen, Chao Dong}
	
	\institute {$^{1}$Shenzhen Key Lab of Computer Vision and Pattern Recognition, SIAT-SenseTime Joint Lab, Shenzhen Institutes of Advanced Technology, Chinese Academy of Sciences \\$^{2}$ University of Chinese Academy of Sciences $^{3}$ University of Macau \\$^{4}$ Shanghai AI Laboratory, Shanghai, China\\
		\email {\{xn.liu, jf.hu1, chao.dong\}@siat.ac.cn, chxy95@gmail.com}}
	
	\maketitle

	\begin{abstract}
		Under-Display Camera (UDC) has been widely exploited to help smartphones realize full-screen displays.
		However, as the screen could inevitably affect the light propagation process, the images captured by the UDC system usually contain flare, haze, blur, and noise.
		Particularly, flare and blur in UDC images could severely deteriorate the user experience in high dynamic range (HDR) scenes.
		In this paper, we propose a new deep model, namely UDC-UNet, to address the UDC image restoration problem with an estimated PSF in HDR scenes.
		%
		%
		Our network consists of three parts, including a U-shape base network to utilize multi-scale information, a condition branch to perform spatially variant modulation, and a kernel branch to
		leverage the prior knowledge of the PSF.
		According to the characteristics of HDR data, we additionally design a tone mapping loss to stabilize network optimization and achieve better visual quality.
		Experimental results show that the proposed UDC-UNet outperforms the state-of-the-art methods in quantitative and qualitative comparisons. 
		Our approach won second place in the UDC image restoration track of the MIPI challenge.
		Codes and models are available at \url{https://github.com/J-FHu/UDCUNet}.\blfootnote{* Equal contributions, $\dagger$ Corresponding author }
	\end{abstract}
	
	\section{Introduction}
	\label{sec:Introduction}
	
\begin{figure}[!t]
		\centering
		\includegraphics[width=1\linewidth]{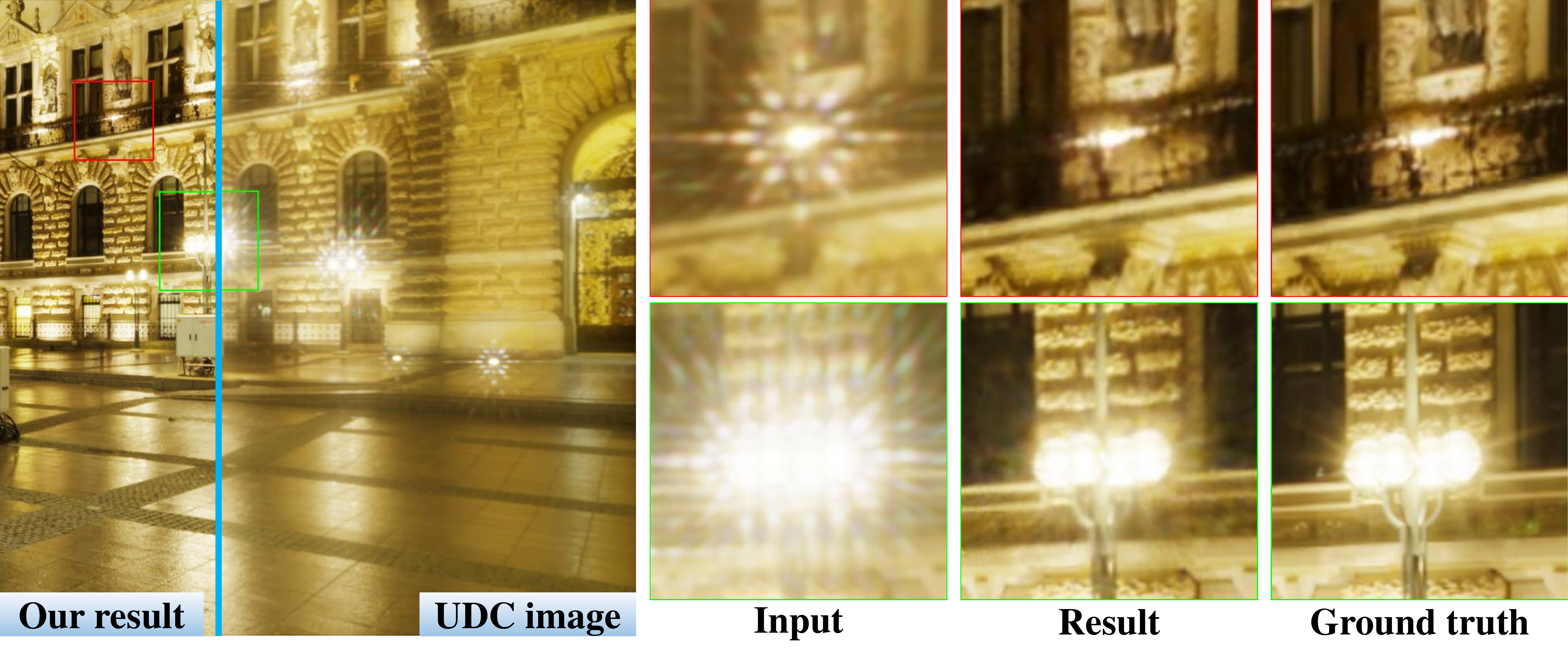}
		\caption{Visual results of under-display camera image restoration. Affected by the UDC imaging mechanism, artifacts including flare, blur, and haze will appear in UDC images. Our method generates visually pleasing results by alleviating the artifacts.} \label{fig:show}
	\end{figure}

	As a new design for smartphones, Under-Display Camera (UDC) systems can provide a bezel-less and notch-free viewing experience, and attracts much attention from the industry. 
	However, it is difficult to preserve the full functionality of the imaging sensor under a display. 
	As the propagation of light in the imaging process is inevitably affected, various forms of optical diffraction and interference would be generated. 
	The image captured by UDC often suffers from noise, flare, haze, and blurry artifacts. 
	Besides, UDC images are often captured under a high dynamic range in real scenes.
	Therefore, severe over-saturation could occur in highlight regions, as shown in Fig. \ref{fig:show}. 
	
	Image restoration task aim to restore the clean image from its degraded version, such as denoising \cite{zhang2017beyond,zhang2018ffdnet,guo2019toward}, deraining \cite{li2016rain,zhu2017joint,li2018recurrent}, deblurring \cite{kupyn2019deblurgan,li2019blind}, super-resolution \cite{srcnn_eccv,rcan,chen2022activating}, and HDR reconstruction \cite{eilertsen2017hdr,chen2021hdrunet}. 
	Similar to these tasks, UDC image restoration aims at reconstructing the degraded image generated by the UDC system. 
	To model the complicated degradation process of the UDC imaging system, existing works \cite{zhou2021image,feng2021removing} propose to utilize a particular diffraction blur kernel, i.e. Point Spread Function (PSF).
	Then UDC image restoration can be regarded as an inversion problem for a measured PSF. 
	Though existing methods have made significant progress in this task, their performance is still limited. 
	
	In this work, we propose UDC-UNet to restore images captured by UDC in HDR scenes. 
	The framework consists of a base network, a condition branch, and a kernel branch. 
	First, we exploit the commonly used U-shape structure to build the base network for utilizing the multi-scale information hierarchically. 
	Then, to achieve the spatially variant modulation for different regions with different exposures, we adopt several spatial feature transform (SFT) layers \cite{wang2018recovering} to build the condition branch as the design in \cite{chen2021hdrunet}. 
	Additionally, the prior knowledge of an accurately measure PSF has been proven to be useful for the restoration process \cite{feng2021removing}. 
	Hence, we also add a kernel branch that uses an estimated PSF to refine the intermediate features. 
	Besides, considering the data characteristic of HDR images, we design a new tone mapping loss to normalize image values into [0,1]. 
	It can not only balance the effects of pixels with different intensities but also stabilize the training process.
	Experimental results show that our method surpasses the state-of-the-art methods in both quantitative performance and visual quality. 
	Our approach can generate visually pleasing results without artifacts even in the over-saturated regions, as shown in Fig. \ref{fig:show}. 
	The proposed UDC-UNet won second place in the MIPI UDC image restoration challenge.
	
	In summary, our contributions are three-fold: 
	
	1) We propose a new deep network, UDC-UNet, to address the UDC image restoration problem. 2) We design a tone mapping loss to address the UDC image restoration problem in HDR scenes. 3) Experiments demonstrate the superiority of our method in quantitative performance and visual quality.
	
	
	\section{Related Work}
	\label{sec:Related Work}

	\subsection{UDC Restoration}
	\label{sec:UDC Restoration}
	
	Under-display camera is a new imaging system that mounts a display screen on top of a traditional digital camera lens.
	While providing a better user experience of viewing and interaction, the UDC system may sacrifice the quality of photography.
	Due to low light transmission rate and diffraction effects in the UDC system, significant quality degradations may appear in UDC images, including flare, haze, blur, and noise.
	%
	%
	Some works~\cite{qin2017evaluation,tang202028} have analyzed the imaging principle of the UDC system and attempted to restore UDC images with deconvolutional-based methods like Wiener Filter~\cite{goldstein1998multistage}.
	Qin et al.~\cite{qin2016see} demonstrated the issue of see-through image blurring of transparent organic light-emitting diodes display.
	Kwon et al.~\cite{kwon2016modeling} paid attention to the analysis of transparent displays.
	Zhou et al.~\cite{zhou2021image} recovered the original signal from the point-spread-function-convoluted image.
	Feng et al.~\cite{feng2021removing} treated the UDC image restoration as a non-blind inverse problem for an accurately measured PSF.
	%
	
	Until now, there is only few works to directly solve the problem of UDC image restoration.
	MCIS~\cite{zhou2021image} modeled the image formation pipeline of UDC and proposed to address the image restoration of UDC using a Deconvolution-based Pipeline (DeP) and data-driven learning-based methods.
	Although this variant of UNet solved many problems of UDC restoration, it lacks consideration of HDR in data generation and PSF measurement.
	A newly proposed structure named DISCNet~\cite{feng2021removing} considered high dynamic range (HDR) and measured the real-world PSF of the UDC system.
	Besides, DISCNet regarded PSF as useful domain knowledge and added a separate branch to fuse its information.
	Experiments have shown the effectiveness of doing so in removing diffraction image artifacts.
	
	However, in HDR scenes, existing methods still could not perform well in both blurring of unsaturated regions and flare in over-saturated areas.
	UDC images in this scenario have different exposures and brightness.
	Previous networks apply the same filter weights across all regions, whose capability are no longer powerful enough in this task.
	Spatial feature transform (SFT) layer is a excellent design provided in SFTGAN~\cite{wang2018recovering} that can achieve affine transformation to the intermediate features and perform a spatially variant modulation.
	A lot of works~\cite{lee2020maskgan,shao2020domain,gu2019blind} have proven the effectiveness of the SFT layer, and especially, HDRUNet~\cite{chen2021hdrunet} demonstrated the effectiveness of the module for HDR reconstruction.
	Thus, we adopt exploit SFT in our network to process the various patterns across the image.
	Additionally, since Feng et al. \cite{feng2021removing} presented that the dynamic convolutions generated by the measured PSF could also benefit the UDC image restoration task in HDR scenes, we utilize this mechanism in our design to further refine the intermediate features.
	
	\subsection{Image Restoration}
	\label{sec:Image Restoration}
	
	Image restoration aims to reconstruct a high-quality image from its degraded version.
	Prior to the deep-learning era, many traditional image processing methods~\cite{calvetti1999iterative,hansen1994regularization,starck2007astronomical,starck2002deconvolution,jauch1994maximum,o1998information} have been proposed for image restoration.
	%
	Since the first deep learning method was successfully applied in low-level vision tasks \cite{dong2014learning}, a flurry of approaches have been developed for various image restoration problems, including super-resolution \cite{dong2016accelerating,rcan,kong2022reflash,chen2022activating}, denoising \cite{zhang2017beyond,zhang2018ffdnet,guo2019toward}, deraining \cite{li2016rain,zhu2017joint}, deblurring \cite{kupyn2019deblurgan,li2019blind}, and HDR reconstruction \cite{eilertsen2017hdr,chen2021hdrunet}. 
	%
	%
	
	As one of the most important task in the field of image restoration, image super-resolution attracts the attention of numerous researchers, and thus a series of methods \cite{dong2016accelerating,rcan,wang2018esrgan,liang2021swinir,li2022blueprint,chen2022activating} have been proposed and made great progress. 
	Denoising is a fundamental task for image restoration, and some good methods have been designed to deal with this problem \cite{zhang2017beyond,zhang2018ffdnet,guo2019toward,lehtinen2018noise2noise,liu2018multi}.
	Deblurring, deraining and HDR reconstruction are also practical problems in the image restoration field. Deblurring methods aim to address the image blur caused by camera shake, object motion, and out-of-focus \cite{kupyn2019deblurgan,li2019blind,gao2019dynamic,zhang2020deblurring,park2020multi,cho2021rethinking}. 
	%
	Deraining aims at removing the rain streaks from the degraded images, which is also a widely studied problem in image restoration \cite{li2016rain,zhu2017joint,fu2017removing,li2018recurrent,ren2019progressive}.
	Reconstructing LDR images to their HDR version can greatly improve the visual quality of images and viewing experience for users. There are also many works focusing on this task \cite{eilertsen2017hdr,chen2021hdrunet,liu2020single,kim2019deep,chen2021new}.
	
	Since UNet \cite{ronneberger2015u} first proposed the U-shape network for biomedical image segmentation, this design is widely used in image restoration task. 
	For example, Cho et al \cite{cho2021rethinking} introduced a U-shape network to deal with image deblurring. 
	Chen et al \cite{eilertsen2017hdr,chen2021hdrunet} also proposed UNet-style networks for single image HDR reconstruction. 
	As U-shape network can better utilize multi-scale information and save computations by deploying multiple down-sampling and up-sampling operations, we also exploit this structure for UDC image restoration task. 
	With the development of large models and the pre-training technique, many works propose to address image restoration problems using Transformer \cite{uformer,liang2021swinir,zamir2022restormer} and try to deal with multiple image restoration task simultaneously \cite{chen2021pre,li2021efficient}. However, there are still few methods for UDC image restoration.
	%
	%
	%
	%
	
	\section{Methodology}
	\label{sec:Methodology}
	
	\subsection{Problem Formulation}
	\label{sec:Problem Formulation}
	
	As mentioned in \cite{feng2021removing}, a real-world UDC image formation model with multiple types of degradations can be formulated as 
	\begin{equation}
		\hat{y}=\phi[C(x * k+n)], 
	\end{equation} \label{degradation}
	where $x$ represents the real HDR scene irradiance. $k$ is the known convolution kernel, commonly referred to as the PSF for this task. 
	``$ * $'' denotes the convolution operation and $n$ represents the camera noise. 
	$C(\cdot)$ denotes the clipping operation for the digital sensor r with limited dynamic range and $\phi(\cdot)$ represents the non-linear tone-mapping function. 
	
	Following \cite{feng2021removing}, we treat this task as a non-blind image restoration problem, where the degraded image $\hat{y}$ and the estimated PSF $k$ are given to restore the clean image. 
	Hence, our proposed network $f$ can be defined as 
	\begin{equation}
		x'=f(\hat{y},k;\theta), 
	\end{equation} \label{formulation}
	where $x'$ denotes the predicted image and $\theta$ represents the network parameters.
	
	
	
	\subsection{Network Structure}
	\label{Network Structure}
	
\begin{figure}[t!]
		\centering
		\includegraphics[width=1\linewidth]{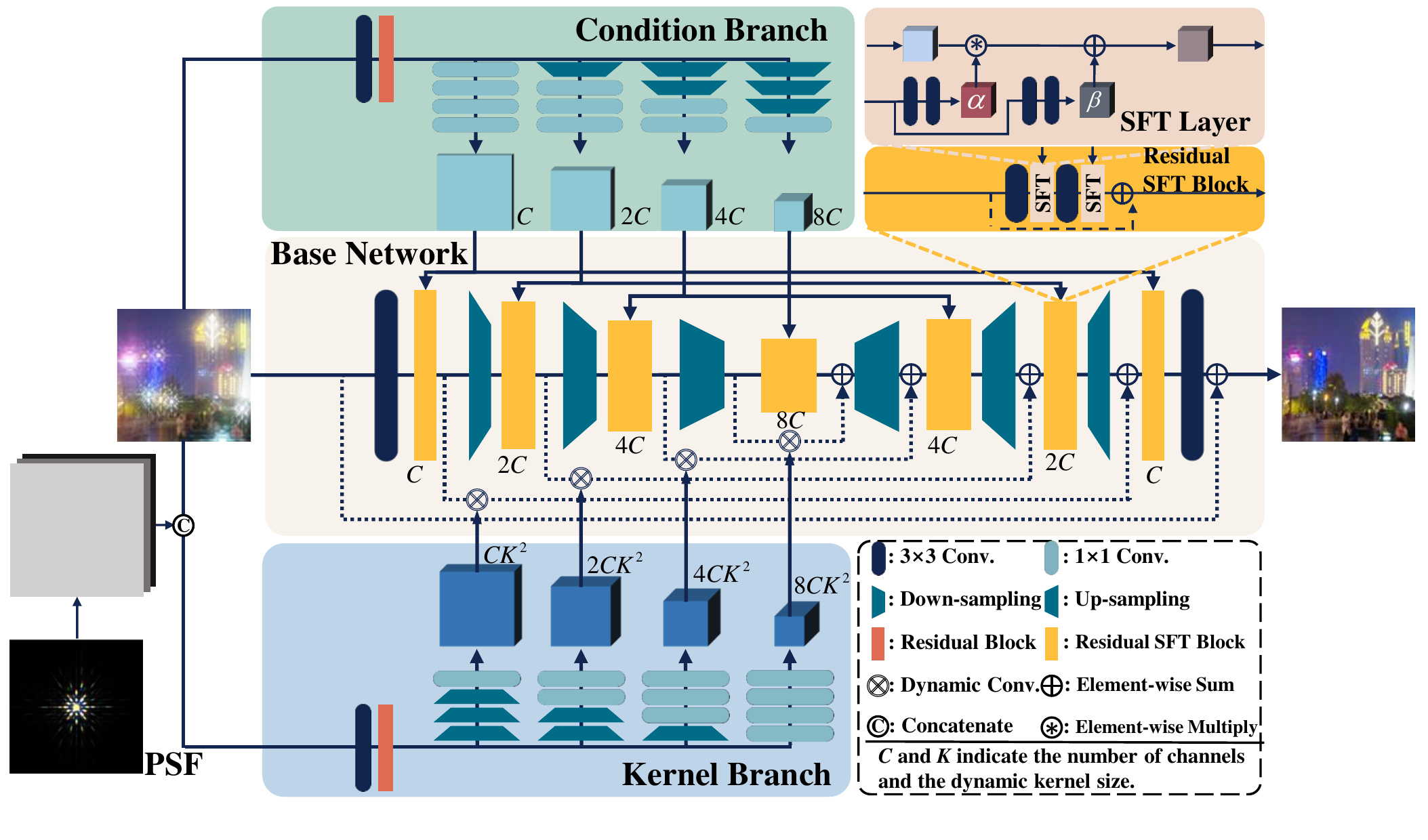}
		\caption{The overall architecture of UDC-UNet. It consists of a U-shape base network, a condition branch, and a kernel branch.}\label{fig:architecture}
	\end{figure}

	The overall architecture of the proposed network consists of three components $–$ a base network, a condition branch, and a kernel branch as shown in Fig. \ref{fig:architecture}, 
	
	\subsubsection{Base Network.}
	\label{sec:Base Network.}
	
	A U-shape structure is adopted to build the base network.
	Its effectiveness has been proven in many image restoration methods~\cite{eilertsen2017hdr,chen2021hdrunet,restormer,uformer}. 
	This kind of structure helps the network make full use of the hierarchical multi-scale information from low-level features to high-level features. 
	The shallow layers in the network can gradually extract the features with a growing receptive field and map the input image to high-dimensional representation. 
	For the deep layers, features can be reconstructed step by step from the encoded representation. 
	Benefiting from the skip connections, shallow features, and deep features can be combined organically. 
	For the basic blocks in the base network, we simply exploit the common residual block with two $3\times3$ convolutions.
	
	\subsubsection{Condition Branch.}
	\label{sec:Condition Branch.}
	
	The key to UDC image restoration in HDR scenes is to deal with the blurring of unsaturated regions and address the flare in over-saturated areas. 
	Traditional convolution filters apply the same filter weights across the whole images, which are inefficient to perform such a spatially variant mapping. 
	HDR-UNet \cite{chen2021hdrunet} proposes to leverage the SFT layer \cite{wang2018recovering} to achieve the spatially variant manipulation for HDR reconstruction with denoising and dequantization. 
	In our approach, we also adopt this module to build the condition branch for adaptively modulating the intermediate features in the base network. 
	
	As demonstrated in Fig. \ref{fig:architecture}, we use the input image to generate conditional maps with different resolutions. 
	Then, these conditions are applied to the features in the base network through the SFT layers. The operation of the SFT layer can be written as 
	\begin{equation}
		SFT(x)=\alpha \odot x+\beta,
	\end{equation}
	where ``$\odot$'' denotes the element-wise multiplication. 
	$x \in\mathbb{R}^{C \times H \times W}$ is the modulated intermediate features in the base network. 
	$\alpha \in\mathbb{R}^{C \times H \times W}$ and $\beta \in\mathbb{R}^{C \times H \times W}$ are two modulation coefficient maps generated by the condition branch. 
	By using such a modulation mechanism, the proposed network can easily perform spatially variant mapping for different regions. 
	Experimental results in Sec. \ref{ablation} shows the effectiveness of this branch for the UDC image restoration task.
	
	
	
	
	
	\subsubsection{Kernel Branch.}
	\label{sec:Kernel Branch.}
	
	Feng et al. \cite{feng2021removing} have demonstrated that the PSF can provide useful prior knowledge for UDC image restoration. 
	Thus, we exploit a similar approach to leverage the PSF to further refine the intermediate features in our base network. 
	Following \cite{gu2019blind,feng2021removing}, we first project the PSF onto a $b$-dimensional vector as the kernel code. 
	Then, we stretch the kernel code into $H\times W\times b$ features and concatenate it with the degraded image of size $H\times W\times C$ to obtain the conditional input of size $H\times W\times (b+C)$. 
	Through the kernel branch, the conditional inputs are mapped into scale-specific kernel features. 
	For the modulated features of $h \times w \times c$, the generated kernel feature is of size $h \times w \times ck^2$. 
	Then, the dynamic convolutions are performed on each pixel as 
	\begin{equation}
		F_{i,j,c} = K_{i,j,c}\cdot M_{i,j,c},
	\end{equation}
	where $K_{i,j,c}$ represents the $k\times k$ filter reshaped from the kernel feature at position $(i,j,c)$ with the size of $1 \times 1 \times k^2$. 
	$M_{i,j,c}$ and $F_{i,j,c}$ denote the $k\times k$ patch centered at $(i,j,c)$ of the modulated features and the element at $(i,j,c)$ of the output feature. `` $\cdot$ '' means the inner product operation. Note that we directly use the measured PSF provided by \cite{feng2021removing} in this paper. 
	
	
	
	\subsection{Loss Function}
	\label{sec:Loss Function}
	
	Traditional loss functions for image restoration task, such as $\ell_{1}$ and $\ell_{2}$ loss, assign the same weight to all pixels regardless of intensity. 
	But for images captured in HDR scenes, such loss function will make the network focus on those pixels with large intensity, resulting in an unstable training process and poor performance. 
	A common practice is to design a specific loss function according to the tone mapping function \cite{chen2021hdrunet,liu2021adnet}. 
	Inspired by these works, we propose a ${Mapping}_{-}\ell_{1}$ loss to optimize the network, which is formulated as
	
	\begin{equation}
		{Mapping}_{-}\ell_{1}(Y, X)=|{Mapping}(Y)-{Mapping}(X)|,
	\end{equation}
	where $Y$ represents the predicted result and $X$ means the corresponding ground truth. For the tone mapping function, ${Mapping}(I)=I/(I+0.25)$.
	
	\section{Experiments}
	
	\subsection{Experimental Setup}
	
	\subsubsection{Dataset.} 
	UDC images and their corresponding ground-truth images in real scenes are currently difficult to obtain. 
	Therefore, Feng \etal in \cite{feng2021removing} propose to use the PSF to simulate the degraded UDC images from the clean images.
	In the MIPI UDC image restoration challenge\footnote{\url{https://codalab.lisn.upsaclay.fr/competitions/4874\#participate}}, 132 HDR panorama images with the spatial size of 8196 $\times$ 4096 from the HDRI Haven dataset\footnote{\url{https://hdrihaven.com/hdris/}} are exploited to generate the synthetic data. 
	This dataset contains images captured in various indoor and outdoor, day and night, nature and urban scenes. 
	The images are first projected to the regular perspective and then cropped into 800 $\times$ 800 image patches. Each of these patches is degraded by using Eqn. (\ref{degradation}) to obtain the simulated UDC images. 
	As the result, 2016 image pairs are available for training. There are also 40 image pairs left as validation set and testing set.
	
	\subsubsection{Implementation Details.} 
	During the training phase, 256 $\times$ 256 patches randomly cropped from the original images are fed into the network. 
	The mini-batch size is set to 32 and the whole network is trained for $6 \times 10^{5}$ iterations. 
	Both the number of residual blocks in the condition branch and kernel branch are set to 2, while the number of residual SFT blocks in the base network is set to [2, 2, 2, 8, 2, 2, 2], respectively. We use the exponential moving average strategy during the training.
	The number of channels $C$ is set to 32 in our UDC-UNet.
	We also provide a small version of our method, UDC-UNet$_S$, by reducing the channel number $C$ to 20.
	The initialization strategy \cite{KN} and Adam optimizer \cite{Adam} are adapted for training. 
	The learning rate is initialized as $2 \times 10^{-4}$ and decayed with a cosine annealing schedule, where $\eta_{min}=1\times 10^{-7}$ and $\eta_{max}=2 \times10^{-4}$. Furthermore, the learning rate is restarted at $[5 \times 10^{4}, 1.5 \times 10^{5}, 3 \times 10^{5}, 4.5 \times 10^{5}]$ iterations. 
	For the evaluation metrics, PSNR, SSIM, and LPIPS are calculated in the tone-mapped images through the tone mapping function ($\operatorname{\textit{Mapping}}(I)=I/(I+0.25)$) provided by the challenge.
	
	\subsection{Ablation Study}
	\label{ablation}
	In this part, we first conduct experiments to explore the effectiveness of the key modules. Then we investigate the influence of different conditional inputs on the kernel branch. Finally, we study the effects of different loss functions. Note that for fast validation, we use the UDC-UNet and set the batch size as 6 for all experiments in this section.
	
	
	\begin{table}[thb]
		\centering
		\renewcommand\arraystretch{1.2}\setlength{\tabcolsep}{10pt}
		\caption{Ablation study of the key modules in UDC-UNet.}\label{Branches}
		\begin{tabular}{lccccc}
			\Xhline{1.2pt}
			Models 		& (a) & (b) & (c) & (d) &(e) (Ours)\\ \hline
			Skip Connections & \ding{56} &\ding{52} &\ding{52} & \ding{52}&\ding{52}\\
			Condition Branch	& \ding{56} &\ding{56} &\ding{56} & \ding{52}&\ding{52} \\
			Kernel Branch	& \ding{56} &\ding{56} & \ding{52}&\ding{56} &\ding{52} \\ \hline
			PSNR(dB) & 42.19 & 44.50 & 44.58 & 45.23 &\textbf{45.37}\\
			SSIM & 0.9884 &0.9897 & 0.9893 &0.9897 &\textbf{0.9898}\\
			LPIPS & 0.0164 & \textbf{0.0155} & 0.0157 &0.0166 & 0.0162\\
			\Xhline{1.2pt}
		\end{tabular}	
	\end{table}
	
	\subsubsection{Effectiveness of Key Modules.}
	As our approach consists of three components -- the base network, the condition branch, and the kernel branch, we conduct ablation experiments to demonstrate the effectiveness of these modules. 
	In addition, we emphasize that the skip connections (all the dotted lines in the base network shown in Fig. \ref{fig:architecture}) in the base network have significant positive effects on performance.
	Our base network without any skip connections and extra branches can obtain the quantitative performance of more than 42dB as shown in Tab. \ref{Branches}, demonstrating the effectiveness of the U-shape structure for this task.
	By comparing models (a) and (b), we can observe that the skip connections bring more than 2dB performance gain. It presents the significance of the skip connection in the base network.
	The models with the condition branch (d,e) outperform the models without this branch (b,c) by more than 0.7dB, illustrating the necessity of the condition branch. 
	The kernel branch also brings a considerable performance gain by comparing the models (c,e) and models (b,d).
	Combining these key modules, our method achieves the highest PSNR and SSIM.
	
	
	\begin{table}[thb]
		\centering
		\renewcommand\arraystretch{1.2}\setlength{\tabcolsep}{10pt}
		\caption{Quantitative results of different conditional inputs for the kernel branch.}\label{Kernel}
		\begin{tabular}{lccc}
			\Xhline{1.2pt}
			Conditional Inputs 		& PSNR(dB)&SSIM &LPIPS\\ \hline
			(a) None	&45.23 &0.9897 &0.0166\\
			(b) Image	&45.17 &0.9896 &\textbf{0.0162}\\
			(c) PSF	& 45.26& 0.9895 & 0.0166\\ 
			(d) Image+PSF &\textbf{45.37} &\textbf{0.9898} &\textbf{0.0162}\\
			
			\Xhline{1.2pt}
		\end{tabular}	
	\end{table}

	\subsubsection{Different Conditional Inputs of the Kernel Branch.}
	An intuitive method for introducing the PSF prior is to directly use the estimated PSF to generate the dynamic filters in the kernel branch. Nonetheless, experiments in \cite{feng2021removing} show that utilizing the combination of the PSF and the degraded image to generate filters can bring more performance improvement. Thus, we also conduct experiments to investigate the different conditional inputs of the kernel branch. As shown in Tab. \ref{Kernel}, using only the degraded image or the PSF does not bring obvious performance gain and even causes a little performance drop. However, when using the image and the PSF simultaneously as the conditional input, the model obtains a nonnegligible performance improvement. 
	
	
	\begin{table}[phb]
		\centering
		\renewcommand\arraystretch{1.2}\setlength{\tabcolsep}{10pt}
		\caption{Quantitative results of our method using different loss functions.}\label{Loss}
		\begin{tabular}{lccc}
			\Xhline{1.2pt}
			Losses 		& PSNR(dB)&SSIM &LPIPS\\ \hline
			$\ell_1$ &41.30 &0.9812 &0.0301 \\
			$\operatorname{\textit{Mapping}}_{-}\ell_2$	&40.19 &0.9838 &0.0238 \\
			$\operatorname{\textit{Mapping}}_{-}\ell_1$	&\textbf{45.37} &\textbf{0.9898} &\textbf{0.0162} \\
			\Xhline{1.2pt}
		\end{tabular}	
	\end{table}
	
	\subsubsection{Effects of Different Loss Functions.}
	We also conduct experiments to show the effects of different loss functions, including the traditional $\ell_1$ loss, the proposed $\operatorname{\textit{Mapping}}_{-}\ell_1$ loss, and $\ell_2$ loss calculated after the same tone mapping function, denoting as the $\operatorname{\textit{Mapping}}_{-}\ell_2$ loss. As depicted in Tab. \ref{Loss}, we can find that our $\operatorname{\textit{Mapping}}_{-}\ell_1$ brings a large quantitative performance improvement compared to the traditional $\ell_1$ loss and also outperforms the $\operatorname{\textit{Mapping}}_{-}\ell_2$ loss by a large margin. The results show that the loss function significantly affects the optimization of the network for this task. 
	
\begin{figure}[t!]
		\centering
		\includegraphics[width=1\linewidth]{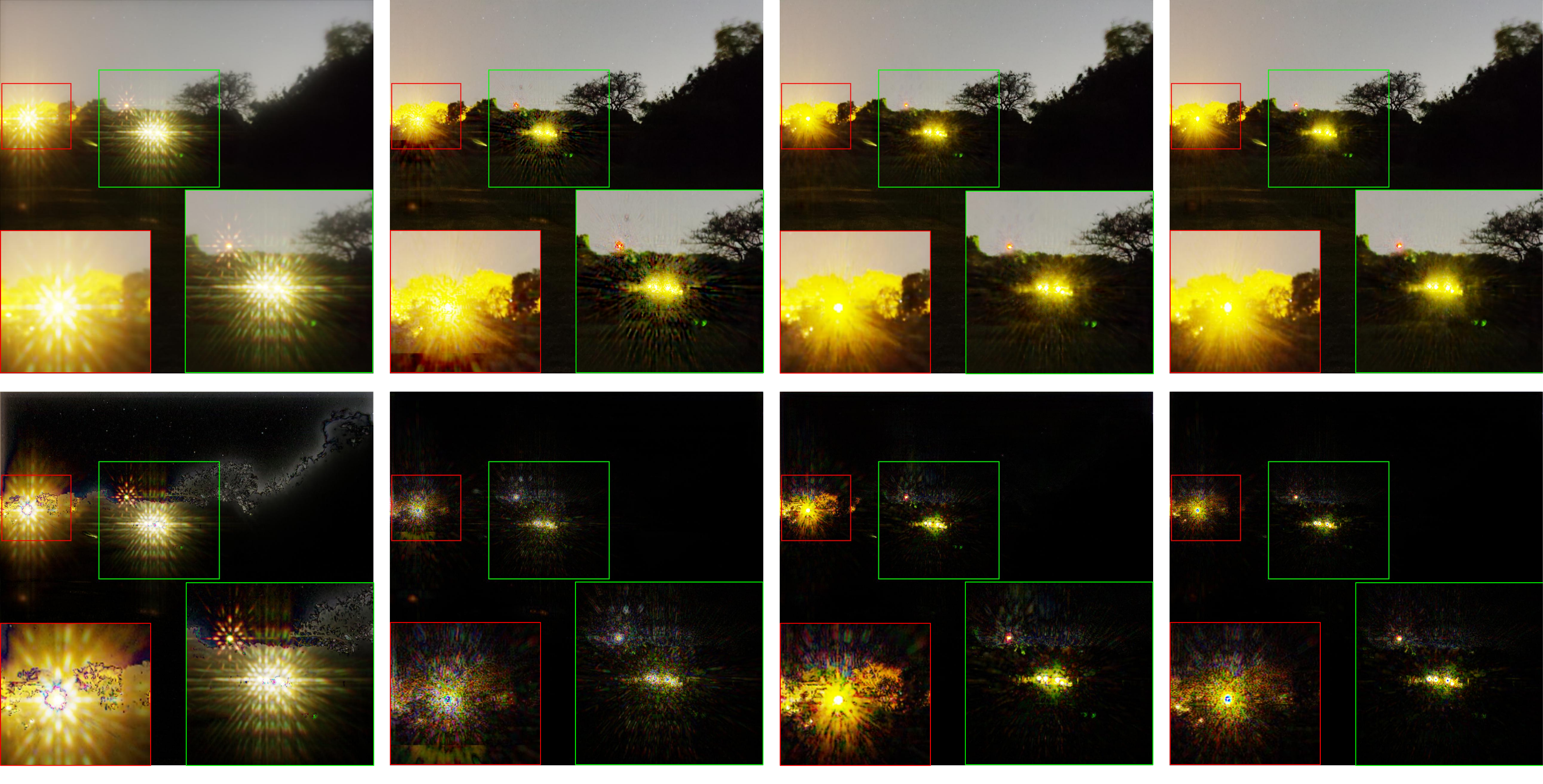}
		\begin{minipage}{ 0.243\linewidth}
			\centering
			{Input\\
				PSNR/SSIM}
		\end{minipage}
		\begin{minipage}{ 0.243\linewidth}
			\centering
			{$\ell_1$\\
				32.09dB/0.9594}
		\end{minipage}
		\begin{minipage}{ 0.243\linewidth}
			\centering
			{$\operatorname{\textit{Mapping}}_{-}\ell_2$\\
				36.47dB/0.9778}
		\end{minipage}
		\begin{minipage}{ 0.243\linewidth}
			\centering
			{$\operatorname{\textit{Mapping}}_{-}\ell_1$\\
				38.18dB/0.9831}
		\end{minipage}
		\vskip -0cm
		\caption{The visual comparisons of models using different loss functions. The bottom row presents the residual maps, E = $\operatorname{\textit{Mapping}}(|Y-X|)$, where $Y$ means the generated output and $X$ indicates the ground truth.}\label{fig:loss}
		\vskip -0.0cm

	\end{figure}

\begin{figure}[h!]
		\begin{minipage}{ 1\linewidth}
			{\includegraphics[width=1\linewidth]{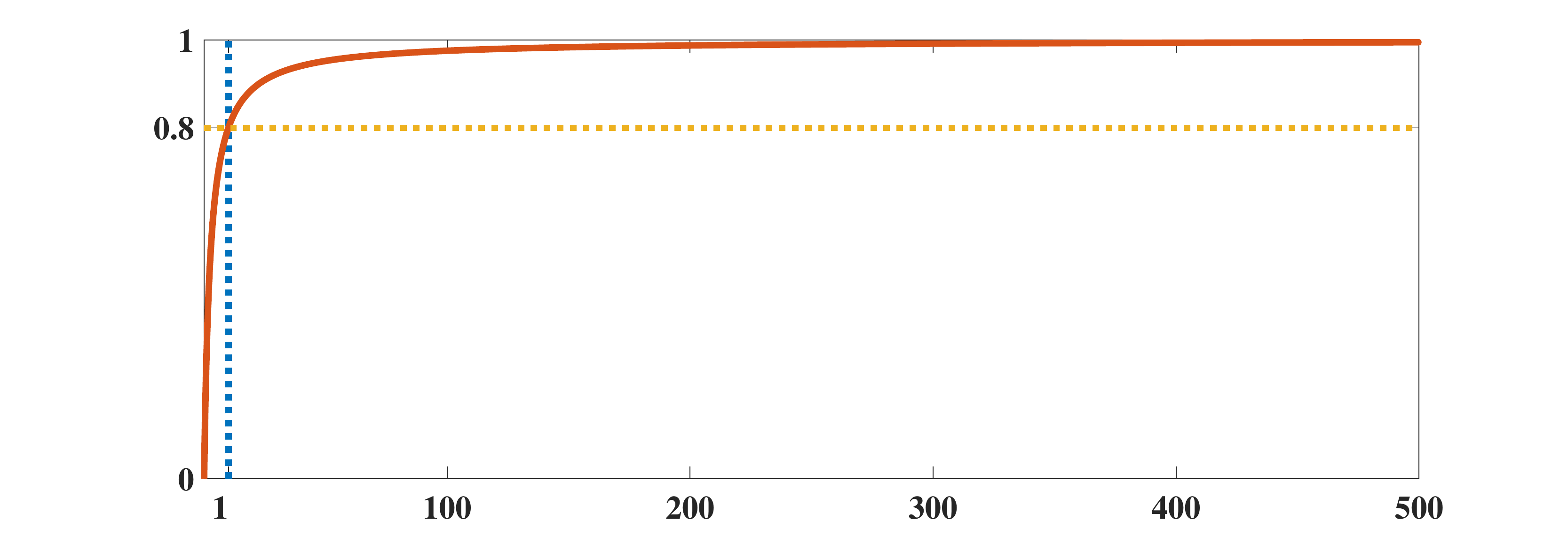}}
			\centering
			{$I\in [0,1] \mapsto \operatorname{\textit{Mapping}}(I) \in [0,0.8]$, $I\in (1,500] \mapsto \operatorname{\textit{Mapping}}(I) \in (0.8,1)$}
		\end{minipage}
			
		\caption{Visualization of the tone mapping function, \ie $\operatorname{\textit{Mapping}}(I)=I/(I+0.25)$.}\label{fig:mapping function}
	\end{figure}

	We also provide the visual comparisons of models using different loss functions, as shown in Fig. \ref{fig:loss}. We can observe that $\ell_1$ loss and $\operatorname{\textit{Mapping}}_{-}\ell_2$ loss make the network focus more on the over-saturated regions, resulting in severe artifacts at these areas. We visualize the tone mapping function in Fig. \ref{fig:mapping function} to further illustrate the effect of our proposed loss. Since the pixel intensities of the used HDR data are in the range [0,500], the large values could significantly affect network optimization and easily make the network focus on the over-saturated regions. After the tone mapping function, the pixel values can be normalized to the range of [0,1). An intuitive manifestation is that values greater than 1 will be compressed to (0.8,1] through the tone mapping function. Hence, the proposed $\operatorname{\textit{Mapping}}_{-}\ell_1$ loss can balance the influence of different regions with different exposures for network optimization.

	\subsection{Comparison with State-of-the-art Methods}
	To evaluate the performance of our UDC-UNet, we conduct experiments on the simulated data and make comparisons with several state-of-the-art methods. 
	Since the task of UDC image restoration is a newly-proposed task, there are not many related works available to compare.
	Therefore, we select three representative methods for comparison, including DISCNet \cite{feng2021removing} for UDC image restoration, Uformer \cite{uformer} for general image restoration tasks, and HDRUNet \cite{chen2021hdrunet} for HDR image reconstruction. 
	Specifically, we retrain the compared methods using the official codes on the same data with the same $\operatorname{\textit{Mapping}}_{-}\ell_1$ loss. Note that we use the small version of Uformer with channel number 32 to save computations. To further demonstrate the effectiveness of our method, we also provide the results of the small version of our method, UDC-UNet$_S$, by reducing the channel number $C$ to 20. 
	
	
	\begin{table}[hbtp]
		\centering
		\renewcommand\arraystretch{1.2}\setlength{\tabcolsep}{6pt}
		\caption{Quantitative comparisons of our method with state-of-the-art approaches. The best values are in \textbf{bold}, and the second best values are underlined.}\label{tab:main_experiment}
			\begin{tabular}{lccccc}
				\Xhline{1.2pt}
				Methods 		& PSNR(dB)&SSIM &LPIPS &Params &GMACs \\ \hline
				Uformer \cite{uformer}	&37.97 &0.9784 &0.0285& 20.0M & 490\\
				HDRUNet \cite{chen2021hdrunet}	&40.23 &0.9832 &0.0240&\textbf{1.7M} & 229\\
				DISCNet \cite{feng2021removing}	&39.89 &0.9864 &0.0152&3.8M & 367\\
				UDC-UNet$_S$ (Ours)	&\underline{45.98} &\underline{0.9913} &\underline{0.0128}&5.7M & \textbf{169}\\ 
				UDC-UNet (Ours)	&\textbf{47.18} &\textbf{0.9927} &\textbf{0.0100} &14.0M & 402\\
				\Xhline{1.2pt}
			\end{tabular}	
		\end{table}

		\subsubsection{Quantitative Results.}
		
		The quantitative results are provided in Tab. \ref{tab:main_experiment}. We can observe that UDC-UNet achieves the best performance and significantly outperforms the compared state-of-the-art methods by more than 7 dB. 
		Considering the computations, we also provide the results of a smaller variant UDC-UNet$_S$. 
		We can observe that UDC-Net$_S$ also obtains considerable performance with relatively fewer parameters and the least computation. The GMACs of all the methods are calculated on the input size of 800 $\times$ 800. Note that due to more downsampling operations, our methods have much fewer GMACs than the compared approaches.
		
		
		\begin{figure}[th!]
			\centering
			\includegraphics[width=1\linewidth]{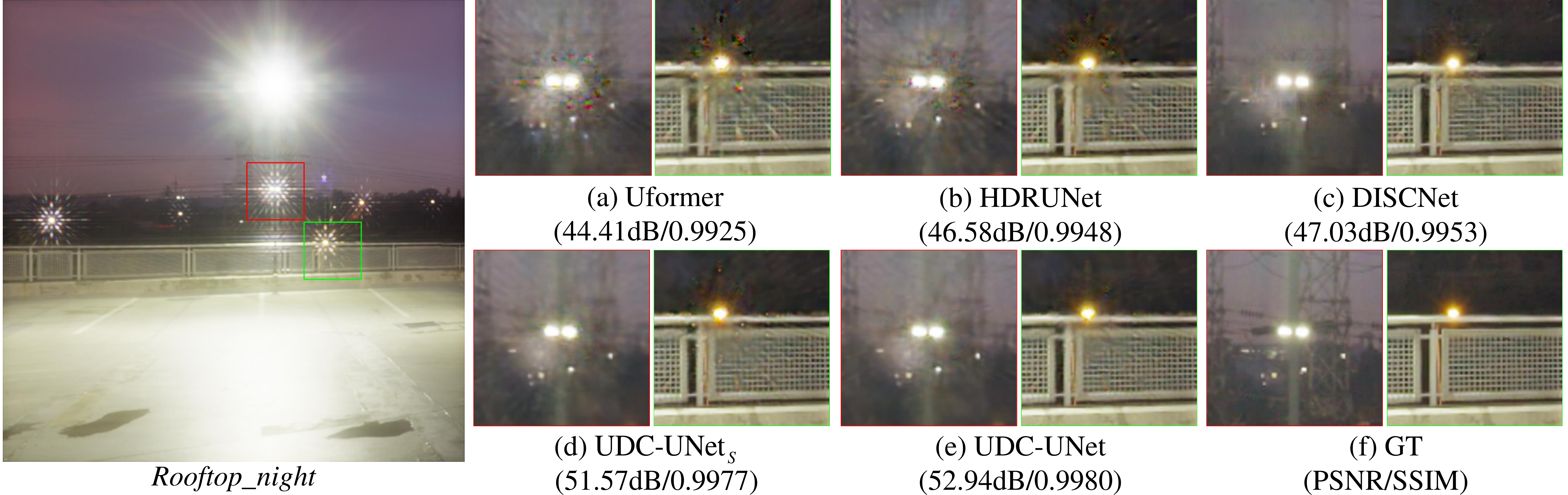}
			\vspace{5pt}
			
			\includegraphics[width=1\linewidth]{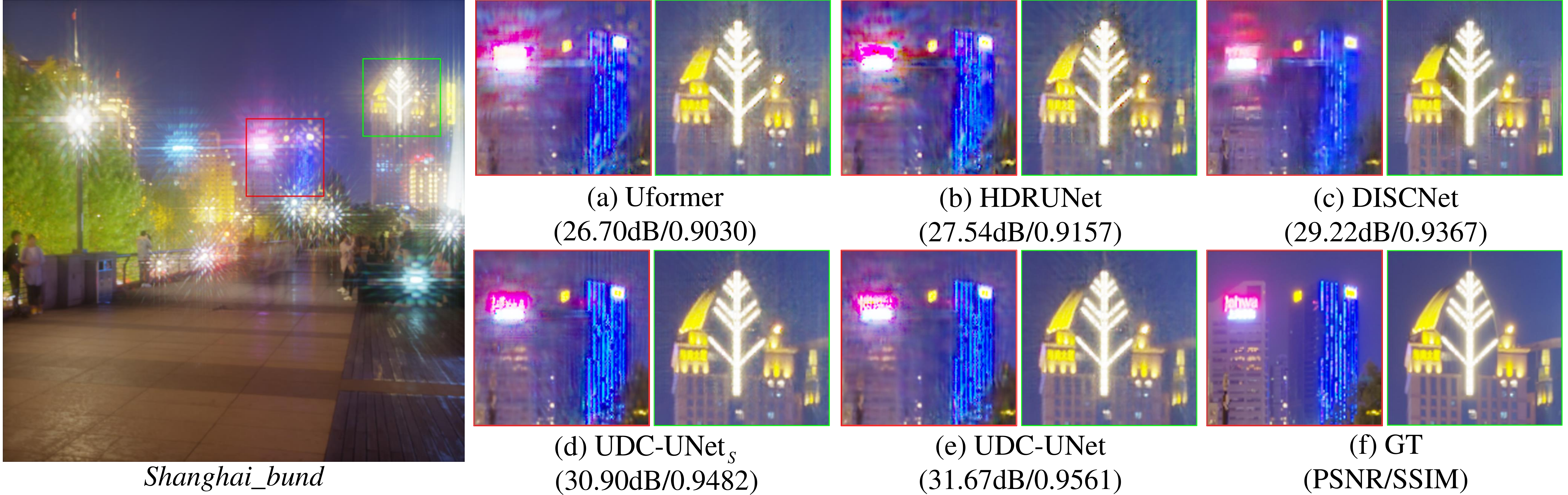}
			\vspace{5pt}
			
			\includegraphics[width=1\linewidth]{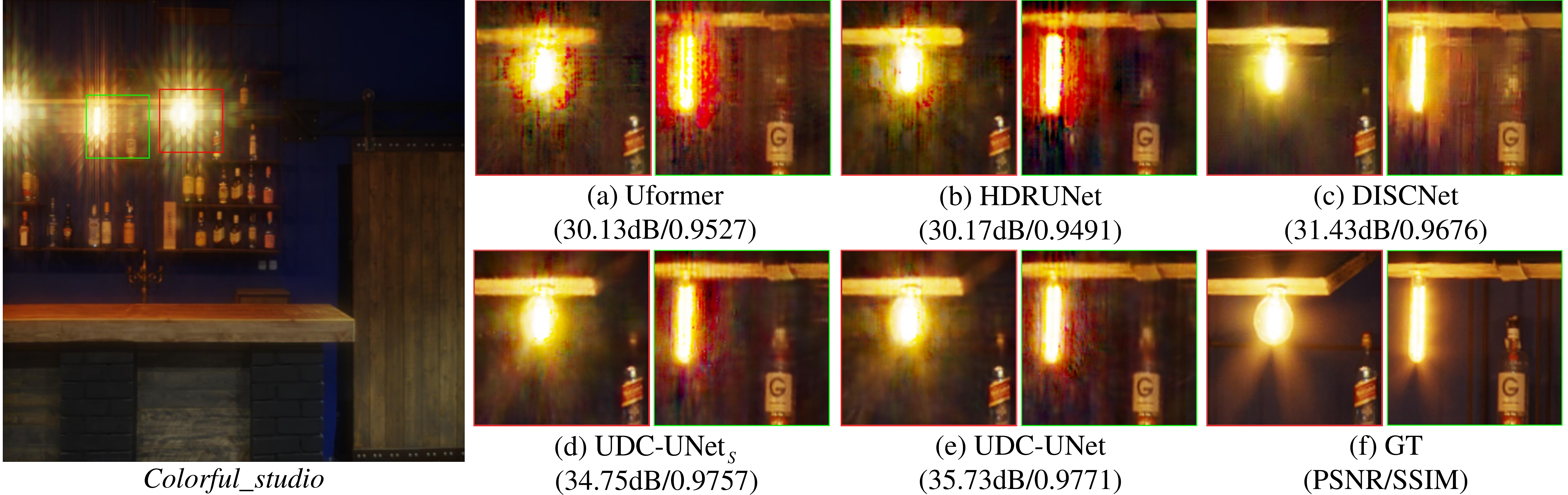}
			\vspace{5pt}
			
			\includegraphics[width=1\linewidth]{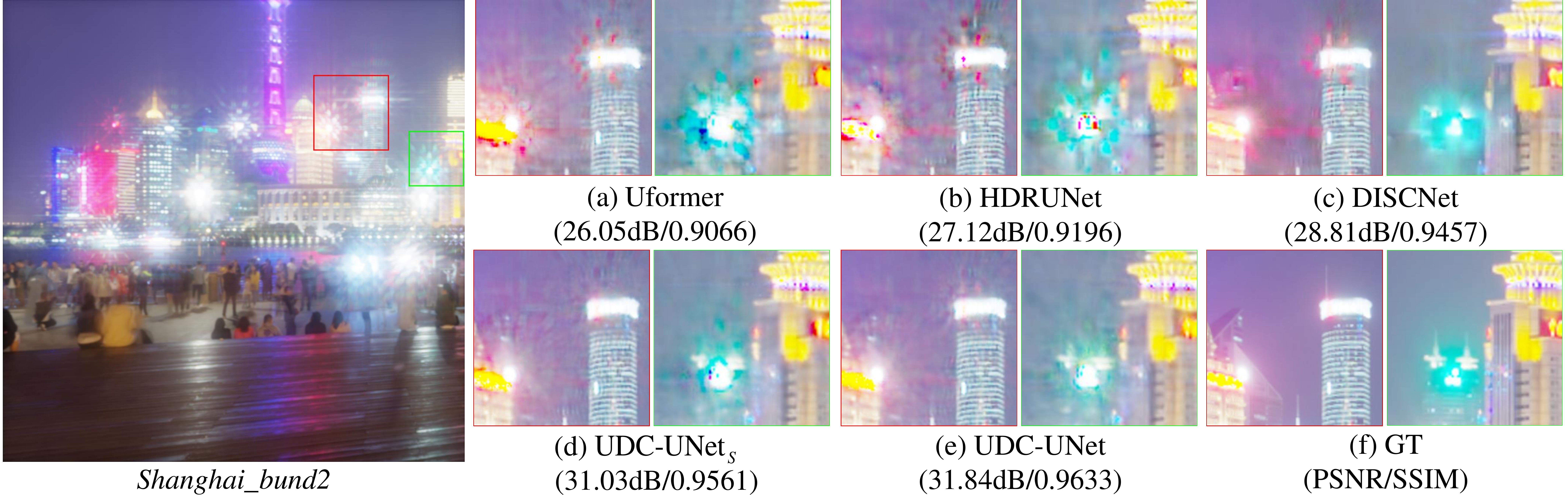}
			\vskip -0cm
			\caption{Visual comparisons of our method with state-of-the-art approaches on the validation set.}\label{fig:main_experiment}
			\vskip -0.0cm
		\end{figure}

		\subsubsection{Visual Comparison.}
		We provide the visual results of our method and state-of-the-art approaches in Fig. \ref{fig:main_experiment}. Compared to other methods, our UDC-UNet obtains visually pleasing results with clean textures and little artifacts, especially on the over-saturated regions. Besides, the results generated by UDC-UNet$_{S}$ also have relatively fewer artifacts compared to other methods. To further demonstrate the superiority of UDC-UNet, we also provide visual comparisons with current models on the testing data. It can be seen that our method obtains better visual quality than the several compared approaches.

		\begin{figure}[tpb!]
			\centering
			\includegraphics[width=1\linewidth]{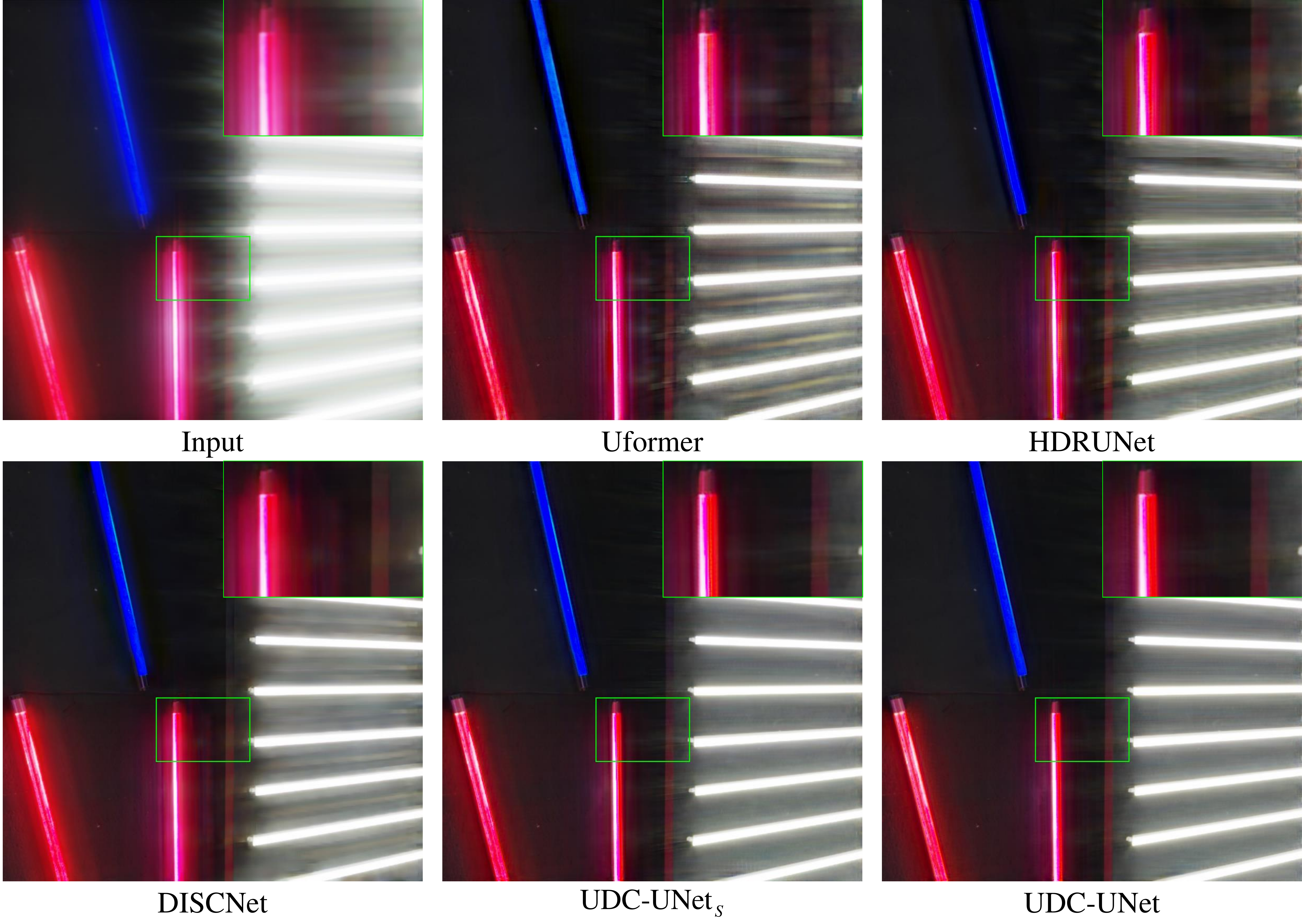}
			\includegraphics[width=1\linewidth]{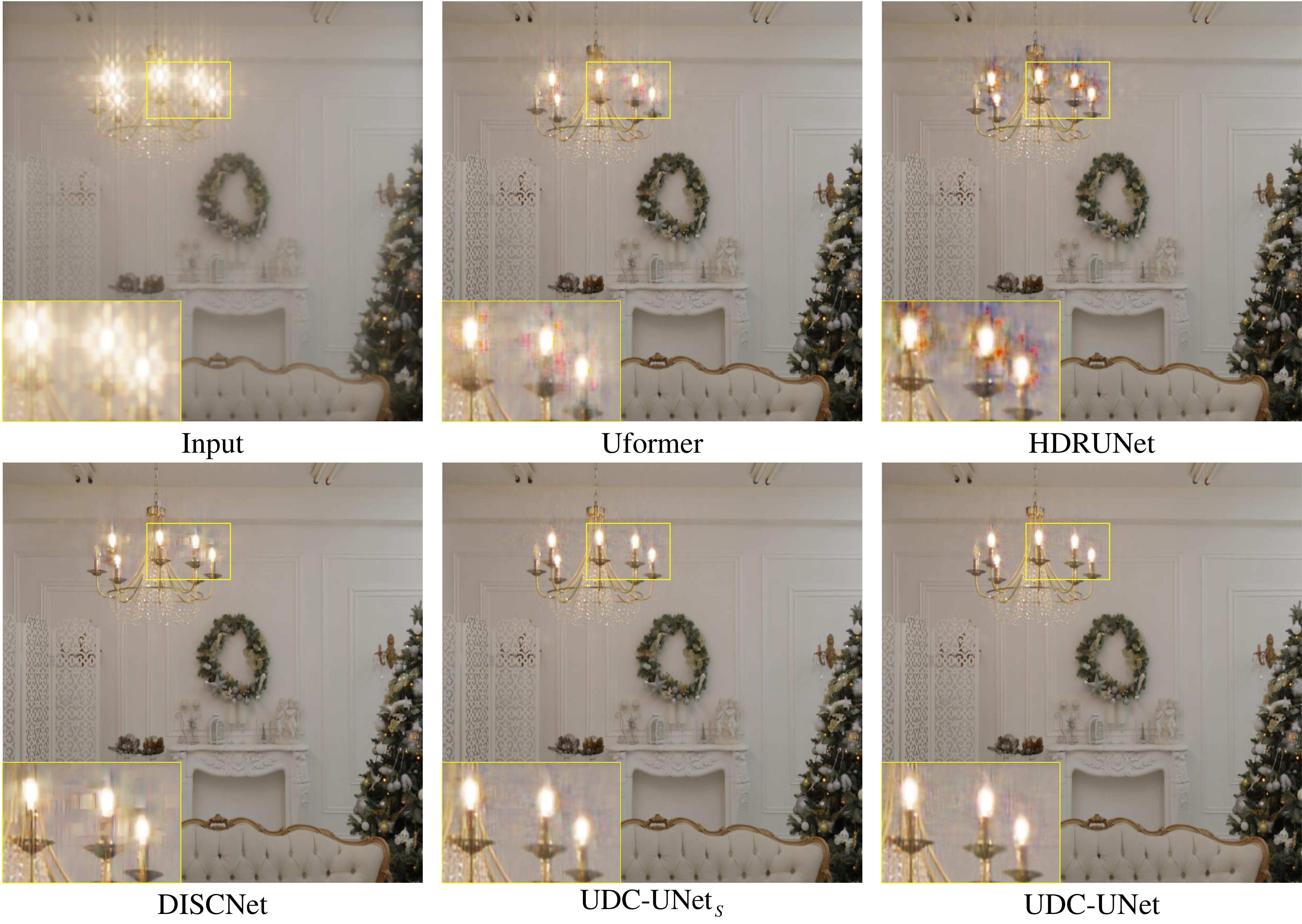}
			\vskip -0cm
			\caption{Visual comparisons of our method with state-of-the-art approaches on the testing set.}\label{fig:report}
			\vskip -0.0cm
		\end{figure}

		\section{Results of the MIPI UDC Image Restoration Challenge}
		
		We participate in the mobile intelligent photography and imaging (MIPI) challenge\footnote{\url{http://mipi-challenge.org/}} and won second place in the UDC image restoration track. The top 5 results are depicted in Tab. \ref{tab:Rank}. 
		
		\begin{table}[thb]
			\centering
			\renewcommand\arraystretch{1.2}\setlength{\tabcolsep}{10pt}
			\caption{Top 5 results of the MIPI UDC image restoration challenge.}\label{tab:Rank}
			\begin{tabular}{lccc}
				\Xhline{1.2pt}
				Teams 		& PSNR(dB)&SSIM &LPIPS\\ \hline
				USTC\_WXYZ	&48.4776 &0.9934 &0.0093 \\
				XPixel Group (ours) 	&47.7821 &0.9913 &0.0122 \\ 
				SRC-B	&46.9227 &0.9929 &0.0098\\ 
				MIALGO	&46.1177 &0.9892 &0.0159 \\ 
				LVGroup\_HFUT	&45.8722 &0.9920 &0.0109 \\ 
				\Xhline{1.2pt}
			\end{tabular}	
		\end{table}

		\section{Conclusions}
		In this paper, we propose UDC-UNet to address UDC image restoration task in HDR scenes. The designed network consists of a U-shape base network to utilize multi-scale information, a condition branch to perform spatially variant modulation, and a kernel branch to leverage the prior knowledge of the measured PSF. In addition, we design a tone mapping loss to balance the effects of pixels with different intensities. Experiments show that our method obtains the best quantitative performance and visual quality compared to state-of-the-art approaches. We participated in the MIPI challenge and won second place in the UDC image restoration track.
		
		\subsubsection{Acknowledgements.} 
		This work is partially supported by  National Natural Science Foundation of China (61906184, U1913210), and the Shanghai Committee of Science and Technology, China (Grant No. 21DZ1100100).
		

		
		
		%
		%
		\bibliographystyle{splncs04}
		\bibliography{egbib}
	\end{document}